\documentstyle[aps,epsfig,multicol]{revtex}

\begin{document}
\title{Cascade production from antikaon induced reactions on
lambda and sigma}
\author{C. H. Li and C. M. Ko}
\address{Cyclotron Institute and Physics Department, Texas A$\&$M University,
College Station, Texas 77843}
\maketitle

\begin{abstract}
Using a gauged flavor SU(3) invariant hadronic Lagrangian, we study 
$\Xi$ production from $\bar K$ induced reactions on $\Lambda$ and 
$\Sigma$ in a coupled-channel approach. Including the four channels 
of $\bar K\Lambda$, $\bar K\Sigma$, $\pi\Xi$, and $\eta\Xi$, 
we solve the Bethe-Salpeter equation in the $K$-matrix
approximation by neglecting the off-shell contribution in the intermediate
states. For the transition potential which drives the higher-order 
contribution, we consider all allowed Born diagrams in the $s$, $t$, 
and $u$ channels. With coupling constants determined from the SU(3) 
symmetry with empirical input, we find that the cross sections 
for the reactions $\bar K\Lambda\to\pi\Xi$, $\bar K\Sigma\to\pi\Xi$, 
$\bar K\Lambda\to\eta\Xi$, and $\bar K\Sigma\to\eta\Xi$ all have values 
of a few mb. In contrast to the results from the lowest-order Born 
approximation, the magnitude of these cross sections is less sensitive 
to the values of the cut-off parameters in the form factors.
From the transition matrix in the coupled-channel approach, we have 
further evaluated the cross sections for the elastic process 
$\bar K\Lambda\to\bar K\Lambda$, $\bar K\Sigma\to\bar K\Sigma$, 
$\pi\Xi\to\pi\Xi$, and $\eta\Xi\to\eta\Xi$ as well as for 
the inelastic processes $\bar K\Lambda\to\bar K\Sigma$ and 
$\pi\Xi\to\eta\Xi$. Implications of the reactions studied here
in $\Xi$ production from relativistic heavy ion collisions are discussed.

\medskip
\noindent PACS number(s): 13.75.Jz
\end{abstract}

\begin{multicols}{2}

%%%%%%%%%%%%%%%%%%%%%%%%%
\section{Introduction}
%%%%%%%%%%%%%%%%%%%%%%%%%%%

Because of the large production rate of strange quarks in a quark-gluon
plasma, enhanced production of hadrons consisting of multistrange 
quarks has been suggested as a possible signal for the quark-gluon 
plasma that is expected to be formed in relativistic heavy ion 
collisions \cite{rafelski}. However, multistrange hadrons can also 
be produced from the hadronic matter that dominates the later stage 
of heavy ion collisions. In particular, the strangeness-exchange 
reactions between antikaons and baryons, such as $\bar K N\to M\Lambda$, 
$\bar K N\to M\Sigma$, $\bar K\Lambda\to M\Xi$, $\bar K\Sigma\to M\Xi$, 
and $\bar K\Xi\to M\Omega$ with $M$ denoting either pion or eta meson,
have been shown in a multiphase transport model \cite{ampt} 
to contribute significantly to the production of multistrange baryons such
as $\Xi$ and $\Omega$ \cite{pal}. Among these reactions, only 
the cross sections for $\bar K N\to\pi\Lambda$ and $\bar K N\to\pi\Sigma$ 
have been measured in experiments \cite{cugnon,exp_dat}.
In Ref.\cite{pal}, the cross sections for other processes are 
determined by assuming that they have the same transition 
matrix elements as that for the empirically measured reaction 
$\bar K N\to\pi\Sigma$. 

To check the validity of the above assumption, we have evaluated 
using the coupled-channel approach the cross sections for $\Xi$
production from the strangeness-exchange reactions induced by 
$\bar K$ on $\Lambda$ and $\Sigma$. The interaction Lagrangians
needed for this study are taken from a gauged flavor SU(3) invariant 
hadronic Lagrangian with empirical hadron masses.  In Refs. 
\cite{su41,su42,su43}, the same Lagrangian based on the SU(4) flavor
symmetry has been used to study reactions involving vector mesons 
and pseudoscalar mesons. Here, we extend it to include the octet
baryons. A Lagrangian similar to ours has been used in 
Refs.\cite{lahiff,kei} to study the empirically 
known cross sections for the reactions $\bar K N\to\pi\Lambda$ and 
$\bar K N\to\pi\Sigma$. Solving the coupled-channel Bethe-Salpeter 
equation \cite{bs_equation} for the transition matrix, it has been
found that the measured cross sections can be reproduced with
appropriate form factors at the interaction vertices. In the present
study, we extend the method of Refs. \cite{kei,k-matrix}, 
which includes only the on-shell part of the propagator in the 
Bethe-Salpeter equation but otherwise satisfies the unitarity 
condition, to study the strangeness-exchange reactions for $\Xi$ production, 
i.e., $\bar K \Lambda\to M\Xi$ and $\bar K \Sigma\to M\Xi$ with $M$ 
denoting either pion or eta meson. 

This paper is organized as follows. In Section \ref{model}, we introduce
the hadronic model used for present study. In particular, the interaction
Lagrangians that are relevant to $\Xi$ production are derived from
a gauged flavor SU(3) invariant Lagrangian with empirical hadron masses. 
Values of the coupling constants are then determined from 
the SU(3) symmetry with empirical input.  
The coupled-channel method is introduced in Section \ref{coupled}. 
To solve the resulting Bethe-Salpeter equation for the transition matrix, 
the lowest-order Born diagrams are evaluated to construct the transition
potential that drives the higher-order contributions, and the $K$-matrix
approximation is used to reduce the Bethe-Salpeter equation to 
a simple form. In Section \ref{details}, details of our
calculations are given. These include the transformation of the
covariant transition matrix between the Dirac spinors to a 
transition matrix between Pauli spinors, and the partial wave
expansion of the transition amplitude.  Results on both the total 
and differential cross sections are given in Section \ref{results}. 
We include not only results on the strangeness-exchange reactions 
for $\Xi$ production but also those on the elastic and inelastic 
processes. We further compare and discuss the results obtained from the 
coupled-channel approach with those from the Born approximation. 
Finally, conclusions are given in Section \ref{conclusions}.  
In the Appendix, we give the explicit expressions of the transition 
amplitudes for all Born diagrams included in present study. 
\newpage

%%%%%%%%%%%%%%%%%%%%%%%%%
\section{The hadronic model}\label{model}
%%%%%%%%%%%%%%%%%%%%%%%%%%

In this section, we introduce the Lagrangian based on the gauged
flavor SU(3) symmetry for describing the interactions between mesons
and baryons. Since the SU(3) symmetry is explicitly broken 
by the appreciable differences in hadron masses, the empirical 
hadron masses are used in the Lagrangian. For the coupling constants, 
we shall determine them using the SU(3) symmetry in terms of 
empirically known coupling constants. 

\subsection{\label{lagrangian}The Lagrangian}

The SU(3) invariant Lagrangian for octet pseudoscalar mesons and
baryons interacting through pseudovector couplings can be written as
\begin{eqnarray}
{\cal L}&=&i{\rm Tr}(\bar B{\partial\mkern-11mu/}B)+{\rm Tr}
[(\partial_\mu P^\dagger)(\partial^\mu P)]+g^\prime \left\{{\rm Tr}[(1-\alpha)
\right.\nonumber\\
&\times&\left.\bar B\gamma^5\gamma^\mu B\partial_\mu P+(1+\alpha)
(\partial_\mu P)\bar B\gamma^5\gamma^\mu B]\right\},
\label{lag}
\end{eqnarray}
where $P$ and $B$ denote, respectively, the $3\times 3$ matrix 
representations of pseudoscalar mesons and baryons, i.e.,  
\begin{eqnarray}
B&=&\left(\begin{array}{ccc}
\frac {\Sigma^0}{\sqrt 2}+\frac{\Lambda}{\sqrt 6} & \Sigma^+ & p \\
\Sigma^- &-\frac{\Sigma^0}{\sqrt 2}+\frac{\Lambda}{\sqrt 6} & n \\
\Xi^- & \Xi^0 & -\sqrt{\frac 2 3}\Lambda 
\end{array}\right), 
\end{eqnarray}
{\footnotesize{
\begin{eqnarray}\label{pseudo}
P&=&\frac {1}{\sqrt 2}\left(\begin{array}{ccc}
\frac {\pi^0} {\sqrt 2} +\frac {\eta_8} {\sqrt 6}+\frac {\eta_0}{\sqrt 3} 
& \pi^+ & K^+\\
\pi^- & -\frac {\pi^0} {\sqrt 2} +\frac {\eta_8} {\sqrt 6} 
+\frac {\eta_0}{\sqrt 3}& K^0\\
K^- & {\bar K}^0 & -\sqrt {\frac 2 3}\eta_8 +\frac {\eta_0}{\sqrt 3}
\end{array}\right).\nonumber\\ 
\end{eqnarray}
}}
In Eq.(\ref{lag}), $g^\prime$ is the universal pseudovector coupling 
of pseudoscalar mesons to baryons, while the parameter $\alpha$ denotes 
the ratio of the $F$ type coupling (${\rm Tr}[(B\bar B+\bar BB)M]$) 
to the $D$ type coupling (${\rm Tr}[(B\bar B-\bar BB)M]$). 

To include the interactions of baryons and pseudoscalar mesons with vector
mesons, we treat vector mesons as gauge particles by replacing the 
partial derivative $\partial_\mu$ in Eq.(\ref{lag}) with
the covariant derivative
\begin{equation}\label{minimal}
D_\mu=\partial_\mu-\frac{i}{2}g[V_\mu,\ ],
\end{equation}
where $V$ denotes the matrix representation of vector mesons, i.e., 
\begin{eqnarray}
V&=&\frac {1}{\sqrt 2}\left(\begin{array}{ccc}
\frac {\rho^0}{\sqrt 2}+\frac{\omega}{\sqrt 2} & \rho^+ & K^{*+} \\
\rho^- &-\frac{\rho^0}{\sqrt 2}+\frac{\omega}{\sqrt 2} & K^{*0} \\
K^{*-} & {\bar K}^{*0} & \phi
\end{array}\right),
\end{eqnarray}
and $g$ denotes the universal coupling of vector mesons with 
baryons and pseudoscalar mesons.

In Eq.(\ref{pseudo}), $\eta_8$ and $\eta_0$ are, respectively, the 
octet and singlet eta mesons, and they are mixtures of physical $\eta$ 
and $\eta^\prime$, i.e.,
\begin{equation}
\eta_8=\cos\theta\ \eta+\sin\theta\ \eta^\prime,\ \eta_0=-\sin\theta\ 
\eta+\cos\theta\ \eta^\prime.
\end{equation}
The mixing angle $\theta$ is not well determined and has a value 
between $-10^{\circ}$ and $-23^{\circ}$ \cite{mixing}.  In the present study, 
we choose the mixing angle to be $-23^{\circ}$ but will study how our 
results change when it is taken to be $-10^{\circ}$.  Also, we shall 
consider only the $\eta$ meson in the present study as the $\eta^\prime$ 
meson is more massive and will not contribute significantly to $\Xi$ 
production from the strangeness-exchange reactions in a hadronic matter. 

\subsection{Interaction Lagrangians}

Expanding the Lagrangian in Eq. (\ref{lag}) using the explicit
matrix representations of $P$, $V$, and $B$, we obtain the following
interaction Lagrangians that are relevant to $\Xi$ production in 
strangeness-exchange reactions:
\begin{eqnarray}
{\cal L}_{\rho\pi\pi}&=&g_{\rho\pi\pi}\vec\pi\cdot\partial_\mu\vec\pi\times
         \vec\rho^\mu,\nonumber\\ 
{\cal L}_{\rho KK}&=&-ig_{\rho KK}
         (\bar K\vec\tau\partial_\mu K-\partial_\mu\bar K\vec\tau K)
         \cdot\vec\rho^\mu,\nonumber\\   
{\cal L}_{\omega KK}&=&-ig_{\omega KK}
         (\bar K\partial_\mu K-\partial_\mu\bar KK)\omega^\mu,\nonumber\\     
{\cal L}_{\phi KK}&=&-ig_{\phi KK}
         (\bar K\partial_\mu K-\partial_\mu\bar K K)\phi^\mu,\nonumber\\ 
{\cal L}_{K^*K\pi}&=&-ig_{K^*K\pi}(\bar K\vec\tau K^{*\mu}\cdot 
         \partial_\mu\vec\pi-\partial_\mu \bar K\vec\tau K^{*\mu}\cdot
         \vec\pi)\nonumber\\
         &+&{\rm H.c},\nonumber\\
{\cal L}_{K^*K\eta}&=&-ig_{K^*K\eta}(\bar K K^{*\mu}\partial_\mu\eta
         -\partial_\mu\bar K K^{*\mu}\eta)+{\rm H.c},\nonumber\\
{\cal L}_{\pi\Lambda\Sigma}&=&\frac{f_{\pi\Lambda\Sigma}}{m_\pi}
	 \bar\Lambda\gamma^5\gamma^\mu\vec\Sigma\cdot\partial_\mu\vec\pi
         +{\rm H.c},\nonumber\\
{\cal L}_{\pi\Sigma\Sigma}&=&i\frac{f_{\pi\Sigma\Sigma}}{m_\pi}
	 \vec{\bar\Sigma}\gamma^5\gamma^\mu\cdot\vec\Sigma\times
         \partial_\mu\vec\pi,\nonumber\\
{\cal L}_{\pi\Xi\Xi}&=&\frac {f_{\pi\Xi\Xi}}{m_\pi} 
	 \bar\Xi\gamma^5\gamma^\mu\vec\tau\Xi
         \cdot\partial_\mu\vec\pi,\nonumber\\
{\cal L}_{\eta\Lambda\Lambda}&=&\frac{f_{\eta\Lambda\Lambda}}
        {m_\eta}\bar\Lambda\gamma^5\gamma^\mu\Lambda\partial_\mu\eta,
	 \nonumber\\
{\cal L}_{\eta\Sigma\Sigma}&=&\frac{f_{\eta\Sigma\Sigma}}{m_\eta}
	 \bar\Sigma\gamma^5\gamma^\mu\Sigma
         \partial_\mu\eta,\nonumber\\
{\cal L}_{\eta\Xi\Xi}&=&\frac{f_{\eta\Xi\Xi}}{m_\eta} 
	 \bar\Xi\gamma^5\gamma^\mu\Xi\partial_\mu\eta\nonumber\\
{\cal L}_{KN\Lambda}&=&\frac{f_{KN\Lambda}}{m_K}
         \bar N\gamma^5\gamma^\mu\Lambda\partial_\mu K+{\rm H.c.},
         \nonumber\\
{\cal L}_{KN\Sigma}&=&\frac{f_{KN\Sigma}}{m_K}
         \bar N\gamma^5\gamma^\mu\vec\tau\cdot\vec\Sigma\partial_\mu K
	 +{\rm H.c.},\nonumber\\
{\cal L}_{K\Lambda\Xi}&=&\frac{f_{K\Lambda\Xi}}{m_K}
	 \bar\Xi\gamma^5\gamma^\mu\Lambda\partial_\mu {\bar K}^T
	 +{\rm H.c},\nonumber\\
{\cal L}_{K\Sigma\Xi}&=&\frac {f_{K\Sigma\Xi}}{m_K}
	 \bar \Xi\gamma^5\gamma^\mu\vec\tau \cdot\vec\Sigma
         \partial_\mu {\bar K}^T+{\rm H.c},\nonumber\\
{\cal L}_{\rho\Sigma\Sigma}&=&ig_{\rho\Sigma\Sigma} 
         \vec{\bar\Sigma}\gamma^\mu\cdot\vec\Sigma\times
         \vec\rho^\mu,\nonumber\\
{\cal L}_{\rho\Xi\Xi}&=&g_{\rho\Xi\Xi}\bar\Xi\gamma^\mu\vec\tau\Xi
	 \cdot\vec\rho^\mu\nonumber\\
{\cal L}_{K^*\Lambda\Xi}&=&g_{K^*\Lambda\Xi}\bar\Xi\gamma^\mu\Lambda
         {\bar K^{*T}_\mu}+{\rm H.c},\nonumber\\
{\cal L}_{K^*\Sigma\Xi}&=&g_{K^*\Sigma\Xi}\bar\Xi\gamma^\mu\vec\tau
         \cdot\vec\Sigma {\bar K^{*T}_\mu}+{\rm H.c},
\label{interaction}
\end{eqnarray}
In the above, $\vec\tau$ are Pauli matrices; $\vec\pi$, $\vec\rho$, and 
$\vec\Sigma$ denote the pion, rho meson, and sigma hyperon isospin 
triplets, respectively, with the usual phase convention of defining
the positively charged states with an extra minus sign with respect 
to the isospin state $|I=1,I_z=1>$. Furthermore, 
$\bar K=(K^-,\bar K^0)$, and $\bar K^*=(K^{*-},\bar K^{*0})$ 
denote the pseudoscalar and vector antikaon isospin doublets,
respectively, and $\bar\Xi=(\bar\Xi^-,\bar\Xi^0)$ is the cascade 
hyperon isospin doublet. 

In terms of the SU(3) coupling constants $g$, $g^\prime$, $\alpha$, and 
the mixing angle $\theta$, the hadronic coupling constants introduced 
in Eq.(\ref{interaction}) can be expressed as 
\begin{eqnarray}
&&g_{\rho\pi\pi}=\frac {g}{4},\ g_{\rho KK}=-\frac g 4,
    \ g_{\omega KK}=-\frac {g}{4},\ g_{\phi KK}=\frac {g}{2\sqrt 2},\nonumber\\
&&g_{K^{*}K\pi}=\frac{g}{4},\  g_{K^*K\eta}=\frac{\sqrt 3}{4}\cos\theta\ g,
    \nonumber\\
&&g_{\rho\Sigma\Sigma}=-\frac g 2,\ g_{\rho\Xi\Xi}=-\frac g 4,
    \ g_{K^* \Lambda \Xi}=\frac {\sqrt 3}{4}g,
    \ g_{K^* \Sigma\Xi}=\frac {g}{4},\nonumber\\
&&\frac{f_{\pi\Lambda\Sigma}}{m_\pi}=\frac {1}{\sqrt 3}g^\prime,
    \ \frac{f_{\pi\Sigma\Sigma}}{m_\pi}=-\alpha g^\prime,
    \ \frac{f_{\pi\Xi\Xi}}{m_\pi}=\frac{1-\alpha}{2} g^\prime,\nonumber\\ 
&&\frac{f_{\eta\Lambda\Lambda}}{m_\eta}=-\left(\frac {\cos\theta}
    {\sqrt 3}+\sqrt{\frac 2 3}\sin\theta\right)g^\prime,\nonumber\\
&&\frac{f_{\eta\Sigma\Sigma}}{m_\eta}=\left(\frac{\cos\theta}
    {\sqrt 3}-\sqrt{\frac 2 3}\sin\theta\right)g^\prime,\nonumber\\
&&\frac{f_{\eta\Xi\Xi}}{m_\eta}
      =-\left(\frac{3 \alpha+1}{2 \sqrt 3}\cos\theta+\sqrt{\frac 2 3}
	\sin\theta\right)g^\prime,\nonumber\\
&&\frac{f_{KN\Lambda}}{m_K}=-\frac{3\alpha+1}{2\sqrt 3}g^\prime,
  \ \frac{f_{KN\Sigma}}{m_K}=\frac{1-\alpha}{2}g^\prime,\nonumber\\
&&\frac{f_{K\Lambda\Xi}}{m_K}=\frac {3\alpha-1}{2\sqrt 3} g^\prime,
  \ \frac{f_{K\Sigma\Xi}}{m_K}=\frac {1+\alpha} {2}g^\prime.
\label{parameters}
\end{eqnarray}

The above hadronic coupling constants can be related to the coupling constants 
between pion and rho meson with nucleon, which are defined by
\begin{eqnarray}
{\cal L}_{\pi NN}&=&\frac{f_{\pi NN}}{m_\pi}\bar N\gamma^5\gamma^\mu
\vec\tau N\cdot\partial_\mu \vec\pi,\nonumber\\ 
{\cal L}_{\rho NN}&=&g_{\rho NN}\bar N\gamma^\mu\vec\tau N\cdot\vec\rho_\mu,
\end{eqnarray}
with
\begin{equation}
\frac{f_{\pi NN}}{m_\pi}=\frac {1+\alpha}{2}g^\prime,~~
g_{\rho NN}=\frac g 4.
\end{equation}
From the empirical values $f_{\pi NN}=1.00$, 
$g_{\rho NN}=3.25$ \cite{tensor}, and $D/(D+F)=1/(1+\alpha)=0.64$ \cite{DF},
we obtain $g^\prime=9.2$ GeV$^{-1}$, $g=13.0$, and $\alpha=0.56$. 
Substituting these parameters into Eq. (\ref{parameters}) gives us the 
hadronic coupling constants in the interaction Lagrangians. Their values 
are given in Table I.

\medskip
\begin{center}
\begin{tabular}{c|c||c|c||c|c|c}\hline
     Vertex&$f$&Vertex&$g$&Vertex&$g$&$g^t$\\
     \hline
     $\pi NN$&1.00&$\rho\pi\pi$&3.25&$\rho NN$&3.25&19.8\\
     $\pi\Lambda\Sigma$&0.741&$\rho KK$&-3.25&$\rho\Lambda\Sigma$&0.0&17.9\\
     $\pi\Sigma\Sigma$&-0.722&$\omega KK$&-3.25&$\rho\Sigma\Sigma$&-6.50
      &-18.0\\
     $\pi\Xi\Xi$&0.281&$\phi KK$&4.60&$\rho\Xi\Xi$&-3.25&7.79\\
     $\eta\Lambda\Lambda$&-1.06&$K^*K\pi$&3.25&$\omega\Lambda\Lambda$&0.0
      &10.0\\
     $\eta\Sigma\Sigma$&4.26&$K^{*}K\eta$&5.18&$\omega \Sigma\Sigma$&0.0&32.1\\
     $\eta\Xi\Xi$&-1.98&&&$\phi\Lambda\Lambda$&0.0&28.4\\
     $KN\Lambda$&-3.52&&&$K^*\Lambda\Xi$&5.63&6.52\\
     $KN\Sigma$&0.992&&&$K^*\Sigma\Xi$&3.25&26.4\\
     $K\Lambda\Xi$&0.900&&&&&\\
     $K\Sigma\Xi$&3.54&&&&&\\
     \hline
\end{tabular}
\end{center}
\vspace{0.3cm}
\noindent{TABLE I: Coupling constants used in the present study. They 
are determined from empirically known coupling constants using
relations derived from the SU(3) symmetry.}

\subsection{Tensor interactions}

Besides the vector interactions between vector mesons and baryons
introduced in Eq.(\ref{minimal}) through the minimal substitution, 
there also exist tensor interactions. In the present study, we assume 
that the tensor interactions are SU(3) invariant and have both $D$ and 
$F$ type as the interactions between pseudoscalar
mesons and baryons. The vector meson and baryon interaction 
Lagrangian can then be written as 
\begin{eqnarray}\label{tensor}
{\cal L}_{VBB}&=&g{\rm Tr}(\bar B\gamma^\mu V_\mu B)
+\frac {g^t}{2m}{\rm Tr}\left[(1-\alpha)\bar B\sigma^{\mu\nu}B
\partial_\mu V_\nu\right.\nonumber\\
&+&\left.(1+\alpha)(\partial_\mu V_\nu)\bar B\sigma^{\mu\nu}B\right],
\end{eqnarray}
where $m$ is the SU(3) degenerate baryon mass and $g^t$ is the universal 
tensor coupling constant. To include the symmetry breaking effect, we 
replace $2m$ in the above equation by the average empirical masses of 
the two baryons at an interaction vertex. One usually introduces the 
parameter $\kappa=g^t/g$ to denote the ratio of tensor to vector couplings.

Expanding Eq.(\ref{tensor}) using the matrix representations of
the vector mesons and baryons, we obtain the following 
relations among tensor coupling constants
\begin{eqnarray}\label{tensor1}
\frac{g^t_{\rho\Lambda\Sigma}}{m_\Lambda+m_\Sigma}&=&
   \frac{2}{\sqrt 3(1+\alpha)}\frac{g^t_{\rho NN}}{2m_N}, \nonumber\\
\frac{g^t_{\rho\Sigma\Sigma}}{2m_\Sigma}&=&
   -\frac{2\alpha}{1+\alpha}\frac{g^t_{\rho NN}}{2m_N},\nonumber\\
\frac{g^t_{\rho\Xi\Xi}}{2m_\Xi}&=&
   \frac{1-\alpha}{1+\alpha}\frac{g^t_{\rho NN}}{2m_N}, \nonumber\\
\frac{g^t_{\omega\Lambda\Lambda}}{2m_\Lambda}&=&
   \frac{2}{3(1+\alpha)}\frac{g^t_{\rho NN}}{2m_N},\nonumber\\
\frac{g^t_{\omega\Sigma\Sigma}}{2m_\Sigma}&=&
   \frac{2}{1+\alpha}\frac{g^t_{\rho NN}}{2m_N},\nonumber\\   
\frac{g^t_{\phi\Lambda\Lambda}}{2m_\Lambda}&=&
   \frac{4\sqrt 2}{3(1+\alpha)}\frac{g^t_{\rho NN}}{2m_N},\nonumber\\   
\frac{g^t_{K^*\Lambda\Xi}}{m_\Lambda+m_\Xi}&=&
   \frac{3\alpha-1}{\sqrt{3}(1+\alpha)}\frac{g^t_{\rho NN}}{2m_N},\nonumber\\
\frac{g^t_{K^*\Sigma\Xi}}{m_\Xi+m_\Sigma}&=&
   \frac{g^t_{\rho NN}}{2m_N}.
\end{eqnarray}

From the empirical value for the rho meson tensor interaction with
nucleons, i.e., $g^t_{\rho NN}=19.8$ \cite{tensor}, we obtain from
Eq.(\ref{tensor1}) the tensor coupling constants shown in Table I.

%%%%%%%%%%%%%%%%%%%%%%%%%%%%%%%%%%%%
\section{The Coupled-channel approach}\label{coupled}
%%%%%%%%%%%%%%%%%%%%%%%%%%%%%%%%%%%%%%%%%%

In Refs. \cite{lahiff,kei}, a coupled-channel approach based on
a Lagrangian similar to the one introduced in Section \ref{model} 
has been used for evaluating the cross sections for
the strangeness-exchange reactions $\bar K N\to\pi\Lambda$ and
$\bar K N\to\pi\Sigma$.  In this paper, we extend the couple-channel 
method to study the reactions $\bar K\Lambda\to\pi\Xi$ and 
$\bar K\Sigma\to\pi\Xi$ for $\Xi$ production. 
We further consider reactions involving an eta meson instead of a pion 
in the final state.

The transition matrix in the present study is a $4\times 4$ matrix 
with matrix elements denoting the transition amplitudes between the
four channels $\bar K\Lambda$, $\bar K\Sigma$, $\pi\Xi$, and $\eta\Xi$.
In the coupled-channel approach, the transition matrix satisfies 
the Bethe-Salpeter equation,
\begin{eqnarray}
T=V+VGT.
\label{BS}
\end{eqnarray}
In the above, $V$ is the transition potential consisting of
all one- and two-particle irreducible connected diagrams, and $G$ 
is the dressed propagator. 

\begin{figure}[h]
\centerline{\epsfig{file=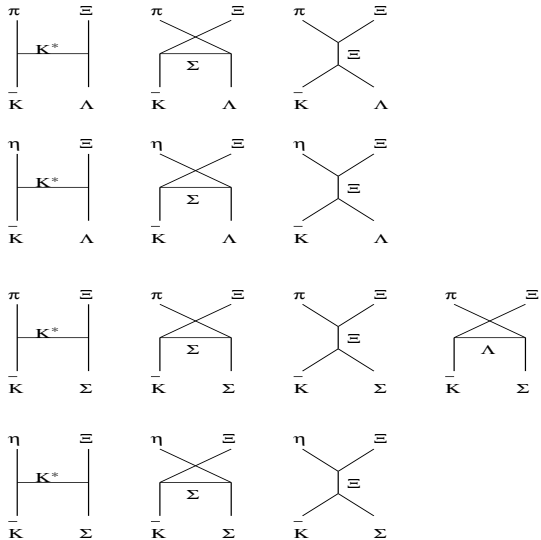,height=2.8in,width=2.8in,angle=0}}
\vspace{0.5cm}
\caption{Born diagrams for $\Xi$ production from strangeness exchange 
reactions.}
\label{born}
\end{figure}

\begin{figure}[h]
\centerline{\epsfig{file=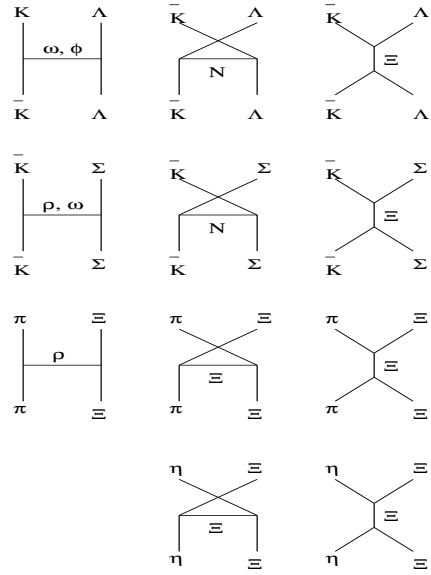,height=3in,width=2.2in,angle=0}}
\vspace{0.5cm}
\caption{Elastic scattering between mesons and hyperons.}
\label{elastic}
\end{figure}

\begin{figure}[h]
\centerline{\epsfig{file=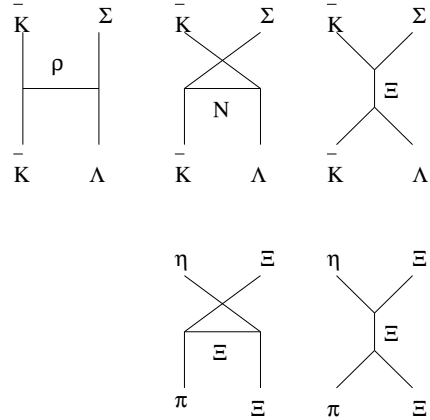,height=2.2in,width=2.2in,angle=0}}
\caption{Inelastic scattering between mesons and hyperons.}
\vspace{0.5cm}
\label{inelastic}
\end{figure}

Following Refs. \cite{lahiff,kei}, we include in the transition potential $V$
only the lowest-order Born diagrams. The diagrams in Fig. \ref{born} 
are for the strangeness-exchange reactions leading to $\Xi$ production. 
Figs. \ref{elastic} and \ref{inelastic} give, respectively, the diagrams
for the elastic and inelastic meson-baryon scattering processes.

The Bethe-Salpeter equation can be rewritten in terms of the imaginary
part of the propagator, $G_{\rm I}$, as 
\begin{equation}
T=K+KG_{\rm I}T,
\end{equation}
where the $K$ matrix satisfies the equation
\begin{equation}
K=V+VG_{\rm R}K,
\end{equation}
with $G_{\rm R}$ denoting the real part of the propagator.

Following the frequently used $K$-matrix approximation, we set $K=V$, i.e., 
neglecting the real part of $G$ or the off-shell contribution in the 
intermediate states. As shown in Ref.\cite{kei}, using the on-shell 
propagator 
\begin{eqnarray}
G_{\rm I}&=&-2i \pi^2\delta(k_m^2-m_m^2)\delta(k_b^2-m_b^2)\nonumber\\ 
&\times&\theta(k_m^0)\theta(k_b^0)({k_b\mkern-16mu/}+m_b),
\label{Gbs}
\end{eqnarray}
with $m_m$ and $m_b$ the masses of intermediate meson and baryon,
respectively, leads to a simple relation between the transition matrix 
$T$ and the transition potential $V$, i.e., 
\begin{equation}
T=\frac{V}{1+iV}.
\label{T-V}
\end{equation}

It can be shown that the $K$-matrix approximation preserves 
the unitarity of the transition matrix \cite{kei}. However, because 
of the neglect of $G_{\rm R}$, the scattering amplitude obtained 
from the $K$-matrix approximation may not have the full 
analytic structure of the exact scattering amplitude.
On the other hand, it has been shown in Ref.\cite{pearce} that 
for pion-nucleon scattering the results from the Bethe-Salpeter
equation are not very sensitive to the approximation used for the 
propagator. 

%%%%%%%%%%%%%%%%%%%%%%%%%%%%%%%%%%%%%%%%%%%%
\section{Calculation details}\label{details}
%%%%%%%%%%%%%%%%%%%%%%%%%%%%%%%%%%%%%%%%%%%

\subsection{Born amplitudes}

Using the interaction Lagrangians in Section \ref{model}, the 
amplitudes for the Born diagrams shown in Figs. \ref{born}, 
\ref{elastic}, and \ref{inelastic} can be straightforwardly 
derived. Aside from isospins, they are given generically by 
\begin{eqnarray}
M_{t}&=&g_1
      \ (q_{i\mu}+q_{f\mu})
      \frac{g^{\mu\nu}-q_t^\mu q_t^\nu/m_{t}^2}{t-m_{t}^2}\nonumber\\
      &\times&\bar u_f\left[(g_2+g^t)\gamma_\nu
      -g^t\frac{p_{i\nu}+p_{f\nu}}{M_i+M_f}\right]u_i,\\
M_{u}&=&\frac{f_i}{m_i}\frac{f_f}{m_f}\frac{1}{u-m_u^2}
      \bar u_f{q\mkern-8mu/}_{i}(m_u-{q\mkern-10mu/}_u){q\mkern-8mu/}_{f}
      u_i,\nonumber\\
M_{s}&=&\frac{f_i}{m_i}\frac{f_f}{m_f}\frac{1}{s-m_s^2}
       \bar u_f{q\mkern-8mu/}_{i}
       (m_s-{q\mkern-10mu/}_s){q\mkern-8mu/}_{f}u_i,
\end{eqnarray}
for the $t$, $u$ and $s$ channels, respectively. In the above, 
$q_i$ and $q_f$ are the momenta of the initial and final mesons
with masses $m_i$ and $m_f$, respectively; $u_i$ is the Dirac spinor
for the initial baryon with mass $M_i$ and momentum $p_i$,
while $u_f$ is that for the final baryon with mass $M_f$ and momentum
$p_f$; and $q_t$, $q_u$ and $q_s$ are the momenta of the intermediate
particles with masses $m_t$, $m_u$ and $m_s$ 
in the three channels. The coupling constants $g_1$ and $g_2$ in the
$t$ channel amplitude denote, respectively, the coupling of the 
exchanged vector meson with pseudoscalar mesons and baryons.
The couplings of the initial and final mesons with baryons in
the $u$ and $s$ channel are given by $f_i$ and $f_f$, respectively.
Explicit expressions including isospins for the amplitudes of all 
diagrams in Figs. \ref{born}, \ref{elastic}, and \ref{inelastic} 
are given in Appendix A.

Because of the finite size of hadrons, form factors are needed at
interaction vertices. All form factors are taken to have the form
\begin{equation}
F({\bf q}^2)=\frac {\Lambda^2}{\Lambda^2+{\bf q}^2}.
\end{equation}
In the above, the momentum ${\bf q}$ is the three momentum transfer 
for $t$ channel diagrams, and the center-of-mass momentum of the 
initial and finial mesons for $u$ and $s$ channel diagrams. The cutoff 
parameter $\Lambda$ is a parameter and will be varied in our study. 
To reduce the number of free parameters, we use the same cutoff
parameter for all vertices.

\subsection{partial wave expansion}

For processes involving pseudoscalar mesons and spin-half baryons as
studied here, the orbital angular momentum is the same for 
the initial and the final state. As shown in Ref. \cite{kei}, the 
relation given in Eq. (\ref{T-V}) between the transition amplitude
and the transition potential obtained in the $K$-matrix
approximation also applies to the transition matrix for a given 
total angular momentum $j$ and orbital angular momentum $l$, i.e., 
\begin{equation}
T^\prime_{jl}=\frac{V^\prime_{jl}}{1+iV^\prime_{jl}},
\end{equation}
where $T^\prime$ and $V^\prime$ are related to $T$ and $V$ by 
\begin{eqnarray}
T(p_i,p_f)=\frac {4\pi }{\sqrt{{\rm p_ip_f}}}T^\prime(p_i,p_f),\nonumber\\
V(p_i,p_f)=\frac {4\pi}{\sqrt{{\rm p_ip_f}}}V^\prime(p_i,p_f),
\end{eqnarray}
with ${\rm p_i}$ and ${\rm p_f}$ the magnitude of
initial and final baryon three momenta in the center-of-mass frame.

The total cross section is then given by 
\begin{equation}\label{cross}
\sigma_{\rm total}=\frac 1{4\pi}\frac {{\rm p_f}}{{\rm p_i}}
\sum_l\left[lT_{l-}^2+(l+1)T_{l+}^2\right],
\end{equation}
where $T_{l\pm}$ denotes $T_{l\pm \frac 1 2,l}$.

To derive $V_{jl}$ from the the Born amplitudes, we use the following
relations given in Refs.\cite{weise,bransden}: 
\begin{equation}
\frac {V_{l\pm}}{4\pi}=\frac 1 2 \int_{-1}^{1}(\tilde AP_l
+\tilde BP_{l\pm 1})dx,
\end{equation}
where $V_{l\pm}=V_{l\pm \frac 1 2,l}$ and
\begin{eqnarray}
\tilde A&=&\frac{\sqrt {(E_i+M_i)(E_f+M_f)}}{8\pi\sqrt s}
[A+B(\sqrt s-\bar M)],\\
\tilde B&=&\frac{\sqrt {(E_i-M_i)(E_f-M_f)}}{8\pi\sqrt s}
[-A+B(\sqrt s+\bar M)].
\end{eqnarray}
In the above, $\bar M=(M_i+M_f)/2$ is the average of the initial 
and final baryon masses; and $E_i$ and $E_f$ 
are, respectively, the initial and final baryon energies.

The spin-independent ($A$) and spin-dependent ($B$) amplitudes 
are defined by 
\begin{equation}
M_{fi}=\bar u(p_f,s_f)\left[A+\frac B 2({q\mkern-10mu/}_i
+{q\mkern-10mu/}_f)\right]u(p_i,s_i),
\end{equation}
and can be evaluated according to  
\begin{eqnarray}
A&=&A_s+A_t+A_u,\nonumber\\
B&=&B_s+B_t+B_u.
\end{eqnarray}
The various terms on the right-hand side of the above equation
are obtained from the Born amplitudes, and aside from isospins 
they are given explicitly by
\begin{eqnarray}
A_s&=&\frac{f_i}{m_i}\frac{f_f}{m_f}\frac{1}{s-m_s^2}\left\{m_s
\left[s-\frac 1 2(M_i^2+M_f^2)\right]\right.\nonumber\\
&-&\left.\frac 1 2(M_i+M_f)(M_iM_f-s)\right\},\\
B_s&=&\frac{f_i}{m_i}\frac{f_f}{m_f}\frac{1}{s-m_s^2}[-m_u(M_i+M_f)
\nonumber\\
&-&s-m_im_f],
\end{eqnarray}
for the $s$ channel Born amplitude, 
\begin{eqnarray}
A_t&=&\frac{g_1}{t-m_t^2}\left[\frac {g_2}{m_t^2}(m_i^2-m_f^2)
(M_i-M_f)\right.\nonumber\\
&-&\left.\frac {g^t} {M_i+M_f}(q_i+q_f)\cdot(p_i+p_f)\right],\\
B_t&=&\frac{2g_1(g_2+g^t)}{t-m_t^2},
\end{eqnarray}
for the $t$ channel Born amplitude, and
\begin{eqnarray}
A_u&=&\frac{f_i}{m_i}\frac{f_f}{m_f}\frac{1}{u-m_u^2}\left\{(m_u+M_f)
\left[s\right.\right.-2E_fE_i\sqrt s\nonumber\\
&+&\left.2p_f\cdot p_i-M_iM_f+\frac 1 2(M_i+M_f)^2\right]
\nonumber\\
&-&\left.\frac 1 2(2q_i\cdot p_f-m_i^2)(M_i-M_f)\right\},\\
B_u&=&\frac{f_i}{m_i}\frac{f_f}{m_f}\frac{1}{u-m_u^2}[(m_u+M_f)(M_i+M_f)
\nonumber\\
&+&m_i^2-2q_i\cdot p_f],
\end{eqnarray}
for the $u$ channel Born amplitude.

%%%%%%%%%%%%%%%%%%%%%%%%%%%%%%%%%%%
\section{Results}\label{results}
%%%%%%%%%%%%%%%%%%%%%%%%%%%%%%%%%%%

In this section, we show both the total and differential cross sections 
obtained from the coupled-channel approach for the strangeness-exchange 
reactions $\bar K\Lambda\to\pi\Xi$, $\bar K\Sigma\to\pi\Xi$, 
$\bar K\Lambda\to\eta\Xi$, and $\bar K\Sigma\to\eta\Xi$. These results
will be compared with those obtained from the Born approximation.
We further show the total cross sections for
the elastic processes $\bar K\Lambda\to\bar K\Lambda$,
$\bar K\Sigma\to\bar K\Sigma$, $\pi\Xi\to\pi\Xi$, and 
$\eta\Xi\to\eta\Xi$ as well as for the inelastic processes
$\bar K\Lambda\to\bar K\Sigma$ and $\pi\Xi\to\eta\Xi$. 
All cross sections are obtained from Eq.(\ref{cross}) by averaging 
over initial and summing over final particle spins and isospins.
 
\subsection{Strangeness-exchange reactions}

\begin{figure}[h]
\centerline{\epsfig{file=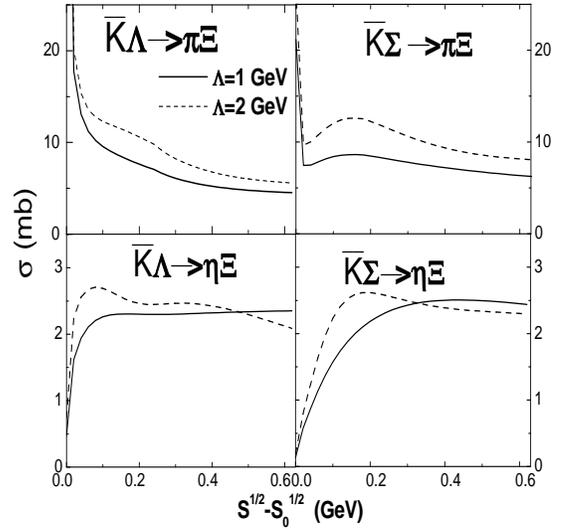,height=2.9in,width=2.9in,angle=270}}
\vspace{0.5cm}
\caption{Cross sections for $\Xi$ production in strangeness exchange 
reactions as functions of center-of-mass energy from the coupled-channel 
approach with cutoff parameters $\Lambda=1$ (solid curves) and 2 GeV 
(dotted curves).} 
\label{result_coupled}
\end{figure}
 
Results on the cross sections for $\Xi$ production from 
the strangeness-exchange reactions $\bar K\Lambda\to\pi\Xi$,
$\bar K\Sigma\to\pi\Xi$, $\bar K\Lambda\to\eta\Xi$, and 
$\bar K\Sigma\to\eta\Xi$ are shown in Fig. \ref{result_coupled}.
These cross sections are evaluated using the coupled-channel approach
with cutoff parameters $\Lambda=1$ and 2 GeV.  It is seen that 
the cross sections do not depend strongly on the value of the cutoff 
parameter. Since the reactions with pions in the final state is
exothermic, their cross sections diverge at threshold. On the 
other hand, the reactions with etas in the final state are endothermic
and have thus vanishing cross section at threshold. For all four
reactions, the cross sections at energies much above threshold 
are a few mb, with those involving pions in the final state
about twice as large as those involving etas in the final state. 

\begin{figure}[h]
\centerline{\epsfig{file=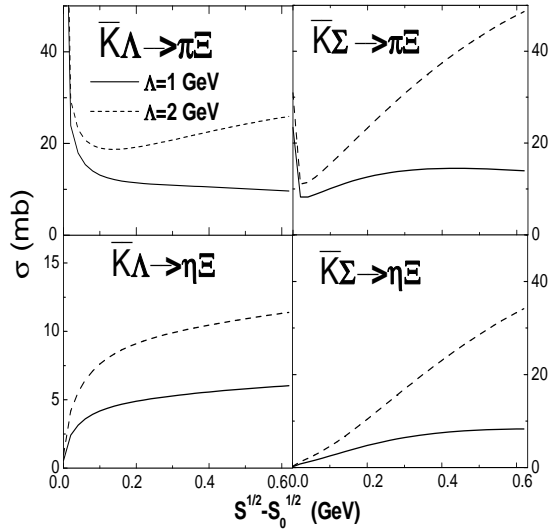,height=2.9in,width=2.9in,angle=270}}
\vspace{0.5cm}
\caption{Cross sections for $\Xi$ production in strangeness exchange 
reactions as functions of center-of-mass energy from the Born
approximation with cutoff parameters $\Lambda=1$ (solid curves) 
and 2 GeV (dotted curves).} 
\label{result_born}
\end{figure}
 
The results from the coupled-channel approach are very different from 
those given by the Born approximation. As shown in Fig. 
\ref{result_born}, the cross sections evaluated using the Born approximation 
have a stronger dependence on the value of the cutoff parameter than
those from the coupled-channel approach.  Furthermore, cross sections
from the Born approximation in general increase with center-of-mass 
energy much faster than those from the coupled-channel approach, 
particularly for the processes $\bar K\Sigma\to\pi\Xi$ and 
$\bar K\Sigma\to\eta\Xi$. As a result, the Born approximation gives 
a much larger cross section than the corresponding one from the 
coupled-channel approach. The difference is large for the
larger cutoff parameter. This is easily seen from Eq.(\ref{T-V}),
as the transition amplitude from the coupled-channel approach
is bounded as the magnitude of the transition potential increases.
Our results thus demonstrate the importance of the unitarity constraint, 
which is satisfied in the coupled-channel approach, on the determination 
of the magnitude for the $\Xi$ production cross sections.

\begin{figure}[h]
\centerline{\epsfig{file=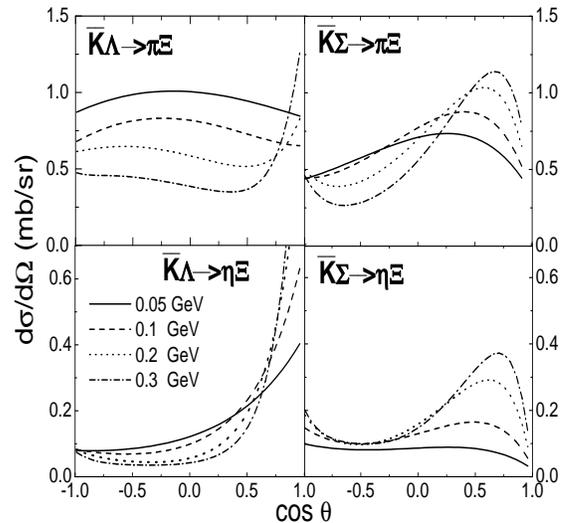,height=2.9in,width=2.9in,angle=270}}
\vspace{0.4cm}
\caption{Differential cross sections for strangeness-exchange reactions
obtained from the coupled-channel approach at different
center-of-mass energies above the threshold.}
\label{angle}
\end{figure}

\begin{figure}[h]
\centerline{\epsfig{file=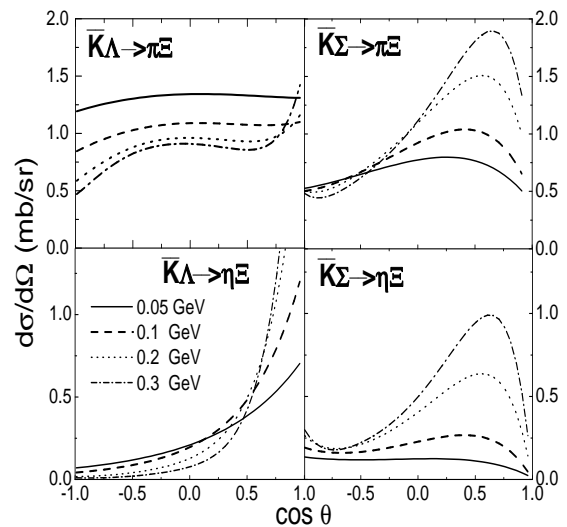,height=2.9in,width=2.9in,angle=270}}
\vspace{0.4cm}
\caption{Differential cross sections for strangeness-exchange reactions
obtained from the Born approximation at different center-of-mass energies 
above the threshold.}
\label{angle_b}
\end{figure}

We have also evaluated the differential cross sections for the
strangeness-exchange reactions, and the results at center-of-mass
energies of 0.05, 0.1, 0.2, and 0.3 GeV above the threshold 
are shown in Fig.\ref {angle}. As expected, the angular distribution
is relatively flat near threshold and becomes more forward peaked 
as the energy increases. For all four processes, the shapes of their
differential cross sections from the coupled-channel approach are, 
however, similar to those obtained from the Born approximation, 
which are shown in Fig. \ref{angle_b}. 

\subsection{Elastic and inelastic scattering processes}

\begin{figure}[h]
\centerline{\epsfig{file=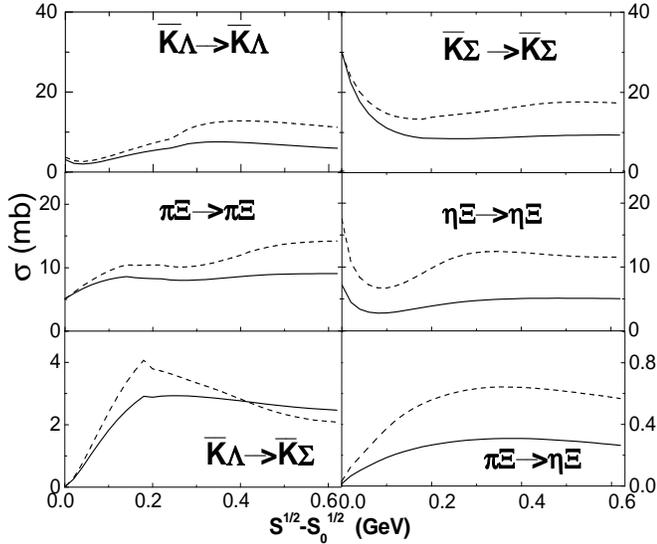,height=3.5in,width=2.9in,angle=270}}
\vspace{0.5cm}
\caption{Cross sections for elastic and inelastic processes 
from the coupled-channel approach for cutoff parameters $\Lambda=1$ GeV 
(solid curves) and $\Lambda=2$ GeV (dash curves).}
\label{other_c}
\end{figure}

The transition matrix $T$ also contains information on the cross
sections for the elastic processes $\bar K\Lambda\to\bar K\Lambda$,
$\bar K\Sigma\to\bar K\Sigma$, $\pi\Xi\to\pi\Xi$, and 
$\eta\Xi\to\eta\Xi$ as well as for the inelastic processes
$\bar K\Lambda\to\bar K\Sigma$ and $\pi\Xi\to\eta\Xi$. 
The results are shown in Fig. \ref{other_c}. As in the case
of strangeness-exchange reactions, these cross sections do
not depend strongly on the cutoff parameter. All four elastic 
scattering cross sections are relatively large. 
For inelastic processes, the cross section for
$\bar K\Lambda\to\bar K\Sigma$ is a few mb but it is less than 
one mb for the reaction $\pi\Xi\to\eta\Xi$. 

\begin{figure}[h]
\centerline{\epsfig{file=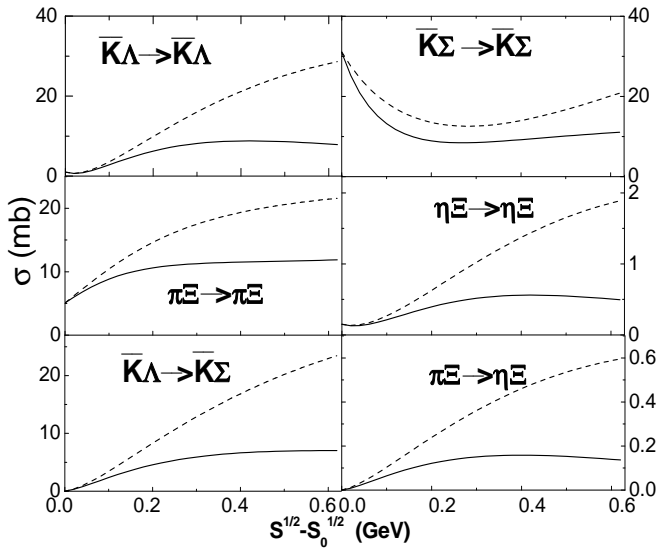,height=3.5in,width=2.9in,angle=270}}
\vspace{0.5cm}
\caption{Cross sections for elastic and inelastic processes obtained
from the Born approximation with cutoff parameters $\Lambda=1$ GeV 
(solid curves) and $\Lambda=2$ GeV (dash curves).}
\label{other_b}
\end{figure}

For comparisons, we show in Fig. \ref{other_b} the cross sections
for the elastic and inelastic processes obtained from the Born
approximation. It is seen that the results depend strongly on the
value of the cutoff parameter. Except for the reaction 
$\eta\Xi\rightarrow\eta\Xi$, the Born approximation 
gives much larger cross sections than the coupled-channel approach
if the cutoff parameter is $\Lambda=2$ GeV. However, with a smaller cutoff
parameter of $\Lambda=1$ GeV, the results from the two approaches 
become comparable.  The cross section for the reaction
$\eta\Xi\rightarrow\eta\Xi$ in the Born approximation is much
smaller than that for the reaction $\pi\Xi\rightarrow\pi\Xi$ due to 
the absence of $t$ channel diagram as shown in Fig. \ref{elastic}.
Comparing the Born cross section for the reaction
$\eta\Xi\rightarrow\eta\Xi$ with that from the coupled-channel approach
shows that the coupled-channel effect increases its cross section
significantly.
 
Another difference between the cross sections from the coupled-channel
approach and those from the Born approximation is that the former
do not have as smooth an energy dependence as the latter. This is
due to the fact that in the coupled-channel approach
the coupling to a different channel in the intermediate states is 
enhanced whenever the energy is above its threshold. 
For example, in the reaction $\pi\Xi\to\pi\Xi$, which has 
the lowest total channel mass, effects of other channels such 
as $\bar K\Lambda$, $\bar K\Sigma$, and $\eta\Xi$ appears successively 
with increasing energy as shown in Fig. \ref{other_c}.

\subsection{Mixing angle dependence}

\begin{figure}[h]
\centerline{\epsfig{file=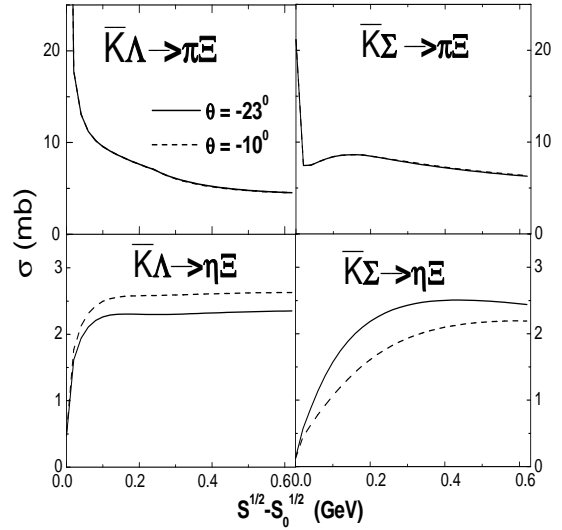,height=2.9in,width=2.9in,angle=270}}
\vspace{0.5cm}
\caption{Cross sections for strangeness-exchange reactions
using two different mixing angles between the singlet and octet
eta mesons.  The cut-off parameter is $\Lambda=1$ GeV in both cases.}
\label{theta_c}
\end{figure}

To see the dependence of our results on the mixing angle between the singlet
and octet eta mesons, we have also calculated the cross sections
for the strangeness-exchange reactions using a mixing angle
of $\theta=-10^{\circ}$ and a cutoff parameter of $\Lambda=1$ GeV.
In Fig. \ref{theta_c}, we compare the results with those using 
$\theta=-23^{\circ}$ and the same cutoff parameter. We find that the cross
sections for the reactions $\bar K\Lambda\to\pi\Xi$ and
$\bar K\Sigma\to\pi\Xi$ are essentially the same for the two 
mixing angles as the eta meson affects their cross sections only
indirectly through the coupled-channel effect. For the reactions 
$\bar K\Lambda\to\eta\Xi$ and $\bar K\Sigma\to\eta\Xi$ involving
the eta meson in the final state, there are slightly variations 
between the cross sections from the two mixing angles.

%%%%%%%%%%%%%%%%%%%%%%%%%%%%%%%%%%%%
\section{conclusions}\label{conclusions}
%%%%%%%%%%%%%%%%%%%%%%%%%%%%%%%%%%%

To evaluate the $\Xi$ production cross section from the strangeness-exchange
reactions between $\bar K$ and hyperons $\Lambda$ and $\Sigma$, we 
have carried out a coupled-channel calculation including the
four channels $\bar K\Lambda$, $\bar K\Sigma$, $\pi\Xi$, and $\eta\Xi$.
The resulting coupled-channel Bethe-Salpeter equation for the
transition matrix is solved in the $K$-matrix approximation, 
i.e., neglecting the real part of the propagator or the off-shell
contribution in the Bethe-Salpeter equation.  

The transition potential, which drives the higher-order effects 
in the Bethe-Salpeter equation, is obtained from the lowest-order Born
diagrams for processes between these channels. The Born diagrams,
which generally include the $s$, $t$, and $u$ channels, are 
evaluated from a gauged SU(3) flavor invariant Lagrangian. 
The symmetry breaking effect is taken into account by using 
empirical masses. For coupling constants in the interaction Lagrangians,
they are determined from empirically known coupling constants 
using relations derived from the SU(3) symmetry. Form factors are 
then introduced at interaction vertices to take into account the finite 
size of hadrons. 

Both total and differential cross sections for the strangeness-exchange
reactions $\bar K\Lambda\to\pi\Xi$, $\bar K\Sigma\to\pi\Xi$, 
$\bar K\Lambda\to\eta\Xi$, and $\bar K\Sigma\to\eta\Xi$ are
evaluated. We find that values of these cross sections 
are not small and are not very dependent on the values of the
cutoff parameters introduced in the form factors.
This is in contrast with the cross sections
evaluated in the Born approximation, which shows a much stronger
dependence on the cutoff parameter. Our results thus demonstrate the
importance of unitarity, which is satisfied by the coupled-channel 
approach in the $K$-matrix approximation but not in the
Born approximation, in evaluating the cross section for
$\Xi$ production from strangeness-exchange reactions. 

As a byproduct of our study, we have also obtained the cross
sections for the elastic process $\bar K\Lambda\to\bar K\Lambda$,
$\bar K\Sigma\to\bar K\Sigma$, $\pi\Xi\to\pi\Xi$, and 
$\eta\Xi\to\eta\Xi$ as well as for the inelastic processes
$\bar K\Lambda\to\bar K\Sigma$ and $\pi\Xi\to\eta\Xi$. 
As for the strangeness-exchange reactions, the cross sections
obtained from the coupled-channel approach are less dependent on
the cutoff parameters than those from the Born approximation. 

The results from our study are useful for relativistic heavy ion 
collisions. Values of the $\Xi$ production cross sections from our
study are not small and are comparable to those used in a recent 
transport model calculations \cite{pal}. In this study, it has been found
that the strangeness-exchange reactions play an important role
in the production of multistrange baryons such as $\Xi$ and $\Omega$
in heavy ion collisions at the SPS energies if one does not assume
that the quark-gluon plasma is formed in the collisions.

Heavy ion collisions at energies below the production threshold
for $\Xi$ in the nucleon-nucleon interaction may be useful
for testing the results obtained in present study. If the heavy
ion collision energy is above the thresholds for $\bar K$, $\Lambda$ 
and $\Sigma$, which are lower than that for $\Xi$, $\Xi$ can still 
be produced via the $\bar K$ induced strangeness-exchange reactions on 
$\Lambda$ and $\Sigma$.  The $\Xi$ yield in heavy ion
collisions at such subthreshold energies is thus sensitive to 
the cross sections for these strangeness-exchange reactions.
A similar idea via the strangeness-exchange reaction $\pi\Lambda\to\bar KN$
\cite{ko,li,cassing} has been found to be important for explaining 
the observed enhancement of $K^-$ production in heavy ion collisions 
at energies below its production threshold in the nucleon-nucleon 
interaction \cite{gsi}.   

For future studies, it would be of interest to extend present
study to $\Omega$ production from the strangeness-exchange reactions
$\bar K\Xi\to\pi\Omega$ and $\bar K\Xi\to\eta\Omega$. This is particular
important as the enhancement of $\Omega$ production in relativistic 
heavy ion collisions is even larger than that for $\Xi$ \cite{sps}.
Knowledge of these cross sections is thus needed to determine if these 
strangeness-exchange reactions also contribute significantly to
$\Omega$ production in relativistic heavy ion collisions.
Since $\Omega$ belongs to the decuplet representation of SU(3),
both the interaction Lagrangians and the partial wave expansion
involved in the calculation will be different from those for
$\Xi$, which belongs to the octet representation of SU(3).

It will also be of interest to improve the results from the
coupled-channel approach by including the off-shell effects due to 
the real part of the propagator in the Bethe-Salpeter equation,
which has been neglected in the present study.

Finally, there also exists other approach to the reactions
$\bar KN\to\pi\Lambda$ and $\bar KN\to\pi\Sigma$ based on the
$SU(3)\times SU(3)$ chiral Lagrangian \cite{lutz}. 
Since chiral symmetry imposes additional constrains on meson-baryon
scattering amplitudes at low-energies, it is important to 
see how the results obtained from present approach
are modified by chiral symmetry. 

\section*{acknowledgments}

We are grateful to Zi-wei Lin and Wei Liu for helpful discussions.
This paper is based on work supported by the National Science 
Foundation under Grant Nos. PHY-9870038 and PHY-0098805, the Welch 
Foundation under Grant No. A-1358, and the Texas Advanced Research 
Program under Grant No. FY99-010366-0081.

%%%%%%%%%%%%%%%%%%%%%%%%%%%%%%%%%%%%%%%%%
\section*{Appendix}
%%%%%%%%%%%%%%%%%%%%%%%%%%%%%%%%%%%%%%%%%%
\setcounter{equation}{0}

In this appendix, we give explicit expressions for 
the amplitudes of all the diagrams in Figs. \ref{born}, 
\ref{elastic}, and \ref{inelastic}.

\noindent{1) $\bar K \Lambda\rightarrow\pi\Xi$:}
\begin{eqnarray}
M_t&=&g_{K^*K\pi}g_{K^*\Lambda \Xi}
      \ \tau^a_{ij}\ (p_{K_i\mu}+p_{\pi^a\mu})\nonumber\\
      &\times&\frac{g^{\mu\nu}-q_t^\mu q_t^\nu/m_{K^*}^2}{t-m_{K^*}^2}
\nonumber\\
      &\times&\bar\Xi_j\left[(1+\kappa)\gamma_\nu
      -\kappa\frac{p_{\Lambda\nu}+p_{\Xi_\nu}}
      {m_\Lambda+m_\Xi}\right]\Lambda,\\
M_u&=&\frac{f_{K \Sigma\Xi}}{m_K}\frac{f_{\pi \Sigma \Lambda}}{m_\pi}
      \ \tau^a_{ij}\ \frac{1}{u-m_\Sigma^2}\nonumber\\
      &\times&\bar\Xi_j{p\mkern-8mu/}_{K_i}(m_\Sigma-{q\mkern-8mu/}_u)
{p\mkern-8mu/}_{\pi^a}\Lambda,\\
M_s&=&\frac{f_{K\Lambda\Xi}}{m_K}\frac{f_{\pi \Xi\Xi}}{m_\pi}
      \ \tau^a_{ij}\ \frac{1}{s-m_\Xi^2}\nonumber\\
      &\times&\bar\Xi_j{p\mkern-8mu/}_{\pi^a}(m_\Xi-{q\mkern-8mu/}_s)
{p\mkern-8mu/}_{K_i}\Lambda.
\end{eqnarray}

\noindent{2) $\bar K \Lambda\rightarrow\eta\Xi$:}
\begin{eqnarray}
M_t&=&g_{K^*K\eta}g_{K^*\Lambda\Xi}
      \ \delta_{ij}\ (p_{K_i\mu}+p_{\eta\mu})\nonumber\\
      &\times&\frac{g^{\mu\nu}-q_t^\mu q_t^\nu/m_{K^*}^2}{t-m_{K^*}^2}
\nonumber\\
      &\times&\bar\Xi_j\left[(1+\kappa)\gamma_\nu
      -\kappa\frac{p_{\Lambda\nu}+p_{\Xi_\nu}}
      {m_\Lambda+m_\Xi}\right]\Lambda,\\
M_u&=&\frac{f_{K\Sigma\Xi}}{m_K}\frac{f_{\eta\Sigma\Lambda}}{m_\eta}
      \ \delta_{ij}\ \frac{1}{u-m_\Lambda^2}\nonumber\\
      &\times&\bar\Xi_j{p\mkern-8mu/}_{K_i}(m_\Lambda-{q\mkern-8mu/}_u)
{p\mkern-8mu/}_\eta\Lambda,\\
M_s&=&\frac{f_{K\Lambda\Xi}}{m_K}\frac{f_{\eta\Xi\Xi}}
      {m_\eta}\ \delta_{ij}\ \frac{1}{s-m_\Xi^2}\nonumber\\
      &\times&\bar\Xi_j{p\mkern-8mu/}_\eta(m_\Xi-{q\mkern-8mu/}_s)
{p\mkern-8mu/}_{K_i}\Lambda.
\end{eqnarray}

\noindent{3) $\bar K \Sigma\rightarrow\pi\Xi$:}
\begin{eqnarray}
M_t&=&g_{K^*K\pi}g_{K^*\Sigma\Xi}
      \ (\tau^a\tau^b)_{ij}\ (p_{K_i\mu}+p_{\pi^a\mu})\nonumber\\
      &\times&\frac{g^{\mu\nu}-q_t^\mu q_t^\nu/m_{K^*}^2}{t-m_{K^*}^2}
\nonumber\\
      &\times&\bar\Xi_j\left[(1+\kappa)\gamma_\nu
      -\kappa\frac{p_{\Sigma\nu}+p_{\Xi_\nu}}
      {m_\Sigma+m_\Xi}\right]\Sigma^b,\\
M_{u1}&=&\frac{f_{K\Sigma\Xi}}{m_K}\frac{f_{\pi\Sigma\Sigma}}{m_\pi}
      \ i\epsilon^{dba}\tau^d _{ij}\ \frac{1}{u-m_\Sigma^2}\nonumber\\
      &\times&\bar \Xi_j{p\mkern-8mu/}_{K_i}(m_\Sigma-{q\mkern-8mu/}_u)
{p\mkern-8mu/}_{\pi^a}\Sigma^b,\\
M_{u2}&=&\frac{f_{K\Lambda\Xi}}{m_K}\frac{f_{\pi\Sigma\Lambda}}{m_\pi}
      \ \delta_{ab}\delta_{ij}\ \frac{1}{u-m_\Lambda^2}\nonumber\\
      &\times&\bar\Xi_j{p\mkern-8mu/}_{K_i}(m_\Lambda-{q\mkern-8mu/}_u)
{p\mkern-8mu/}_{\pi^a}\Sigma^b,\nonumber\\
M_s&=&\frac{f_{K \Sigma\Xi}}{m_K}\frac{f_{\pi\Xi\Xi}}{m_\pi}
      \ (\tau^b\tau^a)_{ij}\ \frac{1}{s-m_\Xi^2}\nonumber\\
      &\times&\bar\Xi_j{p\mkern-8mu/}_{\pi^a}(m_\Xi-{q\mkern-8mu/}_s)
{p\mkern-8mu/}_{K_i}\Sigma^b.
\end{eqnarray}

\noindent{4) $\bar K \Sigma\rightarrow\eta\Xi$:}
\begin{eqnarray}
M_t&=&g_{K^*K\eta}g_{K^*\Sigma\Xi}
      \ \tau^a_{ij}\ (p_{K\mu}+p_{\eta\mu})\nonumber\\
      &\times&\frac{g^{\mu\nu}-q_t^\mu q_t^\nu/m_{K^*}^2}{t-m_{K^*}^2}
\nonumber\\
      &\times&\bar\Xi_j\left[(1+\kappa)\gamma_\nu
      -\kappa\frac{p_{\Sigma\nu}+p_{\Xi_\nu}}
      {m_\Sigma+m_\Xi}\right]\Sigma^a,\\
M_u&=&\frac{f_{K \Sigma\Xi}}{m_K}\frac{f_{\eta \Sigma \Sigma}}{m_\eta}
      \ \tau^a _{ij}\ \frac{1}{u-m_\Sigma^2}\nonumber\\
      &\times&\bar \Xi_j{p\mkern-8mu/}_{K_i}(m_\Sigma-{q\mkern-8mu/}_u)
{p\mkern-8mu/}_\eta\Sigma^a,\\
M_s&=&\frac{f_{K\Sigma\Xi}}{m_K}\frac{f_{\eta \Xi\Xi}}{m_\eta}
      \ \tau^a_{ij}\ \frac{1}{s-m_\Xi^2}\nonumber\\
      &\times&\bar\Xi_j{p\mkern-8mu/}_\eta(m_\Xi-{q\mkern-8mu/}_s)
{p\mkern-8mu/}_{K_i}\Sigma^a.
\end{eqnarray}

\noindent{5) $\bar K \Lambda\rightarrow \bar K\Lambda$:}
\begin{eqnarray}
M_t&=&\left[g_{\omega KK}g^t_{\omega\Lambda\Lambda}
       \frac{g^{\mu\nu}-q_t^\mu q_t^\nu/m_{\omega}^2} {t-M_\omega^2}
       \right.\nonumber\\
      &+&\left.g_{\phi KK}g^t_{\phi\Lambda\Lambda}
       \frac {g^{\mu\nu}-q_t^\mu q_t^\nu/m_{\phi}^2}{t-M_\phi^2}\right]
       \nonumber\\
      &\times&\delta_{ij}\ (p_{K_i\mu}+p_{K_j\mu})\nonumber\\
      &\times&\bar\Lambda\left[\gamma_\nu-\frac{p_{\Lambda\nu}
       +p^\prime_{\Lambda\nu}}{2m_\Lambda}\right]\Lambda,\\
M_u&=&\frac{f_{K N\Lambda}}{m_K}\frac{f_{K N\Lambda}}{m_K}
      \ \delta_{ij}\ \frac{1}{u-m_N^2}\nonumber\\
      &\times&\bar\Lambda{p\mkern-8mu/}_{K_i}(m_N-{q\mkern-8mu/}_u)
      {p\mkern-8mu/}_{\bar K_j}\Lambda,\\
M_s&=&\frac{f_{K \Lambda \Xi}}{m_K}\frac{f_{K\Lambda\Xi}}{m_K}
      \ \delta_{ij}\ \frac{1}{s-m_\Xi^2}\nonumber\\
      &\times&\bar\Lambda{p\mkern-8mu/}_{K_j}(m_\Xi-{q\mkern-8mu/}_s)
{p\mkern-8mu/}_{K_i} \Lambda.
\end{eqnarray}

\noindent{6) $\bar K \Sigma\rightarrow\bar K\Sigma$:}
\begin{eqnarray}
M_t&=&g_{\rho K K}g_{\rho\Sigma\Sigma}
      \ i\epsilon^{abc}\tau^c_{ij}\ (p_{K_i\mu}+p_{K_j\mu})\nonumber\\
      &\times&\frac{g^{\mu\nu}-q_t^\mu q_t^\nu/m_{\rho}^2}{t-m_{\rho}^2}
\nonumber\\
      &\times&\bar\Sigma_a\left[(1+\kappa)\gamma_\nu
      -\kappa\frac{p_{\Sigma^a\nu}+p_{\Sigma^b_\nu}}
      {2 m_\Sigma}\right]\Sigma_b\nonumber\\
      &+&g_{\omega KK}g^t_{\omega\Sigma\Sigma}\delta_{ij}\delta_{ab}
      (p_{K_i\mu}+p_{K_j\mu})\nonumber\\
      &\times&\frac{g^{\mu\nu}-q_t^\mu q_t^\nu/m_{\omega}^2}{t-m_{\omega}^2}
\nonumber\\
      &\times&\bar\Sigma_a\left[\gamma_\nu
      -\frac{p_{\Sigma_a\nu}+p_{\Sigma_b\nu}}
      {2m_\Sigma}\right]\Sigma_b,\\
M_u&=&\frac{f_{K N\Sigma}}{m_K}\frac{f_{K N\Sigma}}{m_K}
      \ (\tau^a\tau^b)_{ij}\ \frac{1}{u-m_N^2}\nonumber\\
      &\times&\bar\Sigma_a{p\mkern-8mu/}_{K_i}(m_N-{q\mkern-8mu/}_u)
{p\mkern-8mu/}_{K_j}\Sigma_b,\\
M_s&=&\frac{f_{K \Sigma\Xi}}{m_K}\frac{f_{K\Sigma\Xi}}{m_K}
      \ (\tau^b\tau^a)_{ij}\ \frac{1}{s-m_\Xi^2}\nonumber\\
      &\times&\bar\Sigma_a{p\mkern-8mu/}_{K_j}(m_\Xi-{q\mkern-8mu/}_s)
{p\mkern-8mu/}_{K_i} \Sigma_b.
\end{eqnarray}

\noindent{7) $\pi \Xi\rightarrow\pi\Xi$:}
\begin{eqnarray}
M_t&=&g_{\rho\pi\pi}g_{\rho\Xi\Xi}
      \ i\epsilon_{abc}\tau^c_{ij}\ (p_{\pi^a\mu}+p_{\pi^b\mu})\nonumber\\
      &\times&\frac{g^{\mu\nu}-q_t^\mu q_t^\nu/m_{\rho}^2}{t-m_{\rho}^2}
\nonumber\\
      &\times&\bar\Xi_j\left[(1+\kappa)\gamma_\nu
      -\kappa\frac{p_{\Xi_f\nu}+p_{\Xi_i\nu}}
      {2 m_\Xi}\right]\Xi_i,\\
M_{u}&=&\frac{f_{\pi\Xi\Xi}}{m_\pi}\frac{f_{\pi\Xi\Xi}}{m_\pi}
      \ (\tau^a\tau^b) _{ij}\ \frac{1}{u-m_\Xi^2}\nonumber\\
      &\times&\bar \Xi_j{p\mkern-8mu/}_{\pi^b}(m_\Xi-{q\mkern-8mu/}_u)
{p\mkern-8mu/}_{\pi^a}\Xi_i,\\
M_s&=&\frac{f_{\pi\Xi\Xi}}{m_\pi}\frac{f_{\pi \Xi\Xi}}{m_\pi}
      \ (\tau^b\tau^a)_{ij}\ \frac{1}{s-m_\Xi^2}\nonumber\\
      &\times&\bar\Xi_j{p\mkern-8mu/}_{\pi^a}(m_\Xi-{q\mkern-8mu/}_s)
{p\mkern-8mu/}_{\pi^b}\Xi_i.
\end{eqnarray}

\noindent{8) $\eta \Xi\rightarrow\eta\Xi$:}
\begin{eqnarray}
M_u&=&\frac{f_{\eta\Xi\Xi}}{m_\eta}\frac{f_{\eta\Xi\Xi}}{m_\eta}
      \ \delta_{ij}\ \frac{1}{u-m_\Xi^2}\nonumber\\
      &\times&\bar \Xi_j{p\mkern-8mu/}_{\eta_i}(m_\Xi-{q\mkern-8mu/}_u)
{p\mkern-8mu/}_{\eta_f}\Xi_i,\\
M_s&=&\frac{f_{\eta\Xi\Xi}}{m_\eta}\frac{f_{\eta \Xi\Xi}}{m_\eta}
      \ \delta_{ij}\ \frac{1}{s-m_\Xi^2}\nonumber\\
      &\times&\bar\Xi_j{p\mkern-8mu/}_{\eta_f}(m_\Xi-{q\mkern-8mu/}_s)
{p\mkern-8mu/}_{\eta_i}\Xi_i.
\end{eqnarray}

\noindent{9) $\bar K \Lambda\rightarrow\bar K\Sigma$:}
\begin{eqnarray}
M_t&=&g_{\rho KK}g^t_{\rho\Sigma\Lambda}
      \ \tau^a_{ij}\ (p_{K_i\mu}+p_{K_j\mu})\nonumber\\
      &\times&\frac{g^{\mu\nu}-q_t^\mu q_t^\nu/m_{\rho}^2}{t-m_{\rho}^2}
\nonumber\\
      &\times&\bar\Sigma_a\left[\gamma_\nu
      -\frac{p_{\Sigma_a\nu}+p_{\Lambda\nu}}
      {m_\Lambda+m_\Sigma}\right]\Lambda,\\
M_u&=&\frac{f_{KN\Sigma}}{m_K}\frac{f_{KN\Lambda}}{m_K}
      \ \tau^a_{ij}\ \frac{1}{u-m_N^2}\nonumber\\
      &\times&\bar \Sigma_a{p\mkern-8mu/}_{K_i}(m_N-{q\mkern-8mu/}_u)
{p\mkern-8mu/}_{K_j}\Lambda,\\
M_s&=&\frac{f_{K\Lambda\Xi}}{m_K}\frac{f_{K\Sigma\Xi}}{m_K}
      \ \tau^a_{ij}\ \frac{1}{s-m_\Xi^2}\nonumber\\
      &\times&\bar\Sigma_a{p\mkern-8mu/}_{K_j}(m_\Xi-{q\mkern-8mu/}_s)
{p\mkern-8mu/}_{K_i}\Lambda.
\end{eqnarray}

\noindent{10) $\pi \Xi\rightarrow\eta\Xi$:}
\begin{eqnarray}
M_u&=&\frac{f_{\pi\Xi\Xi}}{m_\pi}\frac{f_{\eta\Xi\Xi}}{m_\eta}
      \ \tau^a_{ij}\ \frac{1}{u-m_\Xi^2}\nonumber\\
      &\times&\bar \Xi_j{p\mkern-8mu/}_{\pi_a}(m_\Xi-{q\mkern-8mu/}_u)
{p\mkern-8mu/}_{\eta}\Xi_i,\\
M_s&=&\frac{f_{\pi\Xi\Xi}}{m_\pi}\frac{f_{\eta \Xi\Xi}}{m_\eta}
      \ \tau^a_{ij}\ \frac{1}{s-m_\Xi^2}\nonumber\\
      &\times&\bar\Xi_j{p\mkern-8mu/}_{\eta}(m_\Xi-{q\mkern-8mu/}_s)
{p\mkern-8mu/}_{\pi_a}\Xi_i.
\end{eqnarray}

%%%%%%%%%%%%%%%%%%%%%%%%%%%%%%%%%%%%%%%%%%%%%%%%%%%%%
{}

\end{multicols}

\end{document}